# Ensemble Variational Quantum Algorithm for Non-Markovian Quantum Dynamics

Peter L. Walters[1†], Joachim Tsakanikas[2,3†], Fei Wang[1,4,*]

[1]Department of Chemistry and Biochemistry, George Mason University, Fairfax, Virginia 22030, USA
[2]Department of Physics and Astronomy, George Mason University, Fairfax, Virginia 22030, USA
[3]Department of Physics, University of Virginia, Charlottesville, Virginia 22904, USA
[4]Quantum Science and Engineering Center, George Mason University, Fairfax, Virginia 22030, USA
[†]These authors contribute equally to the work

**Abstract**

Many physical and chemical processes in the condensed phase environment exhibit non-Markovian quantum dynamics. As such simulations are challenging on classical computers, we developed a variational quantum algorithm that is capable of simulating non-Markovian dynamics on NISQ devices. We used a quantum system linearly coupled to its harmonic bath as the model Hamiltonian. The non-Markovianity is captured by introducing auxiliary variables from the bath trajectories. With Monte Carlo sampling of the bath degrees of freedom, finite temperature dynamics is produced. We validated the algorithm on the simulator and demonstrated its performance on the IBM quantum device. The framework developed is naturally adapted to any anharmonic bath with non-linear coupling to the system, and is also well suited for simulating spin chain dynamics in a dissipative environment.

## I. Introduction

Simulating open quantum systems dynamics has received increasing attention due to its direct relevance to condensed phase chemistry,[1] many-body physics,[2] quantum biology[3] as well as quantum error correction.[4] Recent advances have uncovered many interesting phenomena in open quantum systems such as non-equilibrium phase transitions,[5,6] entangled state preparation through reservoir engineering,[7,8] and information backflow.[9–11] For studying quantum dynamics in condensed phase chemical environment ranging from solutions[12,13] to molecular aggregates,[14–17] the stereotypical microscopic framework is the spin-boson model[18,19] and its multistate extension.[20,21] The corresponding charge and exciton dynamics often exhibit non-Markovian behavior. Several well-developed numerically accurate methods are available for carrying out such simulations.[22–27] However, the resource requirement on classical computers often grows exponentially with respect to system size and the degree of non-Markovianity.

As first conjectured by Feynman[28] and demonstrated by Lloyd,[29] quantum computer simulations of quantum dynamics can achieve an advantage. A wealth of literature exist for Hamiltonian simulation algorithms,[30–35] with Low and Chuang having realized the optimal query complexity[36]. For open quantum systems, many work has been focused on Markovian dynamics, ranging from theoretical construction of semigroup generators[37–41] to simulating Lindblad dynamics on NISQ devices.[42–46] On the other hand, the development of quantum algorithms for non-Markovian time evolution is still in its infancy. Notable works include the method of locally indivisible maps,[47] the ensembles of Lindblad trajectories,[48] the construction of superoperators from the generalized quantum master equation[49] and the Feynman-Vernon influence functional,[50] and the path-integral-based algorithm.[51]

In this work, we present a non-Markovian quantum algorithm with a NISQ-friendly focus, the time-dependent variational algorithm (TDVA).[52–54] In particularly, we work with the spin-boson model and use the ensemble averaged classical path (EACP)[55–57] to capture the non-Markovian dynamics in a finite temperature bath. The organization of the paper is of the following. In section II, we briefly discuss the EACP approximation using Feynman's path integral framework. In section III, we discuss its implementation in the TDVA setting. In section IV, we present results and discussions. In section V, we offer some concluding remarks.

## II. Ensemble averaged classical path (EACP) approximation

The Hamiltonian for a quantum system linearly coupled to its harmonic bath can be written in the following form:

$$H = \left(\frac{p_s^2}{2m_0} + V_0(s)\right) + \sum_j \left[\frac{P_j^2}{2m_j} + \frac{1}{2}m_j\omega_j^2\left(x_j - \frac{c_j s}{m_j\omega_j^2}\right)^2\right] \qquad (1)$$

where $s$ and $x_j$ denote the system and bath coordinates, respectively, and $c_j$ denotes the system-bath coupling strength. The strength weighted density of modes defines the spectral density:

$$J(\omega) = \frac{\pi}{2}\sum_j \frac{c_j^2}{m_j\omega_j}\delta(\omega - \omega_j) \qquad (2)$$

The bath's influence on the system can be seen as having a time-dependent driving force,

$$H_0 - \sum_j c_j s x_j(t) \qquad (3)$$

where

$$x_j(t) = x_{0,j}\cos\omega_j t + \frac{p_{0,j}}{m_j\omega_j}\sin\omega_j t + \frac{c_j}{m_j\omega_j}\int_0^t dt'\, s(t')\sin\omega_j(t - t') \qquad (4)$$

The non-local memory kernel in the last part of equation (4), termed *back-reaction*[58] (i.e., kicking back by the system), is partially responsible for the non-Markovian dynamics. The other important contributor is from the integration of the phase space variables $x_{0,j}$ and $p_{0,j}$ from the bath. The effects of these two contributions to the non-Markovianity are delineated by Makri using path integral formulation.[55] Below we briefly summarize the main findings that have direct relevance to the current algorithm implementation.

In the absence of the back-reaction, the reduced density matrix (RDM) expresses as

$$\rho_s(s_t^+, s_t^-) = \int \mathcal{D}s^+ \int \mathcal{D}s^- \langle s_0^+|\rho_0(0)|s_0^-\rangle \exp\left\{\frac{i}{\hbar}(S[s^+] - S[s^-])\right\}\tilde{Q}[s^+, s^-] \qquad (5)$$

where

$$\tilde{Q}[s^+, s^-] = \exp\left(\frac{i}{\hbar}\sum_j c_j x_{0,j}\int_0^t dt'\,\Delta s(t')\cos\omega_j t' + \frac{i}{\hbar}\sum_j \frac{c_j p_{0,j}}{m_j\omega_j}\int_0^t dt'\,\Delta s(t')\sin\omega_j t'\right) \qquad (6)$$

describes the influence of the bath on the system's dynamics. The $s^+$ and $s^-$ denote the forward and backward path, respectively, and $\Delta s = s^+ - s^-$. $S[s^+]$ and $S[s^-]$ are the action integrals of the free system, with $\langle s_0^+|\rho_0(0)|s_0^-\rangle$ the initial state. The integral $\int \mathcal{D}s$ sums over all possible paths. Equation (5-6) is time-local in that the propagation of the RDM can be done iteratively with time.

Integrating over the Wigner distribution,

$$W(\boldsymbol{x}_0, \boldsymbol{p}_0) = (\hbar\pi)^{-1}\prod_j \tanh\left(\frac{1}{2}\hbar\omega_j\beta\right)\exp\left[-\tanh\left(\frac{1}{2}\hbar\omega_j\beta\right)\left(\frac{m_j\omega_j x_{0,j}^2}{\hbar} + \frac{p_{0,j}^2}{m_j\omega_j\hbar}\right)\right] \qquad (7)$$

$\tilde{Q}[s^+, s^-]$ turns into

$$\tilde{Q}[s^+, s^-] = \exp\left\{-\sum_j \frac{c_j^2}{2m_j\omega_j\hbar} \coth\left(\frac{1}{2}\hbar\omega_j\beta\right) \int_0^t dt' \int_0^{t'} dt'' \,\Delta s(t')\Delta s(t'') \cos\omega_j(t'-t'')\right\} \quad (8)$$

Immediately from equation (8), the non-Markovian effect is manifested in the *double* time integration. Therefore, tracing out the bath degrees of freedom introduces non-Markovianity. On the other hand, we can employ the reverse by introducing additional degrees of freedom to remove the non-Markovian effect.

When the back-reaction is included, an additional term, $R[s^+, s^-]$, is introduced that further augments the system dynamics,

$$R[s^+, s^-] = \exp\left\{\frac{i}{\hbar}\sum_j \frac{c_j^2}{m_j\omega_j} \int_0^t dt' \int_0^{t'} dt'' \,\Delta s(t')\Delta\bar{s}(t'') \sin\omega_j(t'-t'')\right\} \quad (9)$$

where $\Delta\bar{s} = \frac{1}{2}(s^+ + s^-)$. Together, $Q$ and $R$ form the Feynman-Vernon influence functional[59]

$$IF[s^+, s^-] = Q[s^+, s^-] \times R[s^+, s^-] \quad (10)$$

A notable difference between $Q$ and $R$ is that $Q$ has a temperature dependent term, $\coth\left(\frac{1}{2}\hbar\omega_j\beta\right)$, whereas $R$ does not. By making the analogy with light matter interaction, Makri pointed out[55] that $Q$ is related to the simulated emission and absorption of phonons, and $R$ the spontaneous emission. Immediately following that observation, the zero-point energy effect of this spontaneous emission has diminished effect at high temperature or for low frequency bath. Therefore, with the back-reaction properly omitted,

$$x_j(t) \cong x_{0,j} \cos\omega_j t + \frac{p_{0,j}}{m_j\omega_j} \sin\omega_j t \quad (11)$$

With

$$H(t) = H_0 - \sum_j c_j \, s x_j(t) \quad (12)$$

now local in time, the dynamics can be solved by Markovian propagation with a time-dependent Hamiltonian,

$$i\hbar \frac{\partial}{\partial t}|\psi(t)\rangle = H(t)|\psi(t)\rangle \quad (13)$$

Finally, to achieve the thermal effect, integration over the position and momentum from equation (7) is performed. In practice, Monte Carlo sampling[60] is used for efficient integration of the multidimensional Wigner distribution. The resulting RDM is the ensemble average of the individual RDM originating from a specific $x_0$ and $p_0$. Since the omitting of the back-reaction is analogous to the treating the light as classical, this approach is termed ensemble averaged classical path (EACP).[55–57] It is worth noting that although the back-reaction (the zero-point energy effect) is omitted, the zero-point energy contribution is not completely removed. The Wigner distribution provides the static zero-point energy effect, whereas the back-reaction offers the dynamical one. Another appealing aspect of this approach is that it is not limited to the harmonic bath linearly coupled to the system; the framework can be equally adapted to non-linear coupling and anharmonic environment, provided its initial Wigner distribution is available.[61]

### III. Time-dependent variational algorithm (TDVA)
Three conventional variational principles exit for time-dependent problems, and the McLachlan's variational principle is proved to be numerically stable for the variational quantum algorithm.[62] The McLachlan's variational principle uses the minimization strategy as the following,

$$\delta \left\| \left( \frac{d}{dt} + iH \right) |\psi(\boldsymbol{\theta})\rangle \right\| = 0 \tag{14}$$

where the wavefunction $\psi$ is determined by a set of parameters $\boldsymbol{\theta}$.

In the hybrid quantum-classical algorithm, the quantum computer calculates the following[62,63]

$$A_{i,j} = \frac{\partial \langle \psi(\boldsymbol{\theta})|}{\partial \theta_i} \frac{\partial |\psi(\boldsymbol{\theta})\rangle}{\partial \theta_j} \tag{15}$$

$$C_i = \frac{\partial \langle \psi(\boldsymbol{\theta})|}{\partial \theta_i} H |\psi(\boldsymbol{\theta})\rangle \tag{16}$$

The update of the parameters $\boldsymbol{\theta}$ is conducted on the classical computer with the following differential equation,

$$\sum_j A^R_{i,j} \dot{\theta}_j = C^I_i \tag{17}$$

where $R$ and $I$ in the superscript refer the real and the imaginary part, respectively. The real and imaginary part of equation (15) and (16) can be extracted by the modified Hadamard test.[63] In this work, we use RK4 to solve equation (17).

## IV. Results and discussions

In the following, we use spin-boson model to test the algorithm. The Hamiltonian in the EACP limit can be written as

$$H(t) = \hbar \Omega \sigma_x - \left( \sum_j c_j x_j(t) \right) \sigma_z \tag{18}$$

with

$$x_j(t) = x_{0,j} \cos \omega_j t + \frac{p_{0,j}}{m_j \omega_j} \sin \omega_j t \tag{19}$$

We choose the bath to have the Ohmic spectral density

$$J(\omega) = \frac{\pi}{2} \hbar \xi \omega e^{-\omega/\omega_c} \tag{20}$$

where dimensionless $\xi$ is the Kondo parameter that determines the strength of the system-bath coupling, and $\omega_c$ is the cutoff frequency. We use 60 oscillators of different frequencies in the numerical calculation, following the discretization procedure given by Walters et al.[64] We have used the atomic units so that $\hbar=1$.

For a two-level system which requires one qubit, there exits an exact ansatz for the unitary operation that employs the ZXZ decomposition[4]

$$U(\boldsymbol{\theta}) = e^{i\theta_1} R_z(\theta_2) R_x(\theta_3) R_z(\theta_4) \tag{21}$$

The wavefunction then can be parameterized as,

$$|\psi(\boldsymbol{\theta})\rangle = U(\boldsymbol{\theta})|0\rangle \tag{22}$$

With this ansatz, the matrix $A$ and the vector $C$ in equation (17) can be computed, and the exemplar circuits compiled by Qiskit[65] are shown in the appendix. It is worth mentioning that instead of propagating the vectorized density matrix, we are propagating the wavefunction. As a consequence, it automatically saves half of the qubits and the circuits are expected to be short. The dissipative effect is through the average of the bath initial conditions.

Figure 1 shows the population dynamics simulated on the QASM simulator[65] for a particular set of initial conditions $(\boldsymbol{x}_0, \boldsymbol{p}_0)$ drawn from the Wigner distribution (equation 7). In this simulation, we use the parameters $\Omega = 1, \xi = 2, \omega_c = 1.5$, and the inverse temperature $\beta = 1$. Each data point is obtained with 50,000 shots. The quantum algorithm result matches well with the classical benchmark result ("EACP 1IC" in the plot). The classical computing result is obtained by directly solving equation (13). It should be pointed out that the data in figure 1 are without the Monte Carlo averaging.

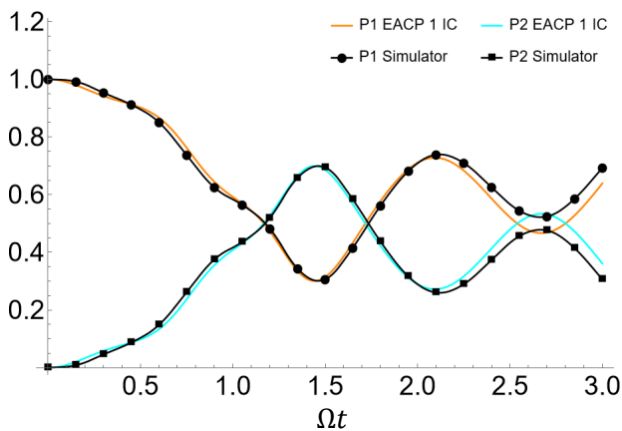

Figure 1. Population dynamics simulated on the simulator for a symmetric two-level system, with one bath initial condition and system initially populated in the reactant state. "P1" and "P2" label the population dynamics of the reactant and product, respectively. Parameter: $\Omega = 1, \xi = 2, \omega_c = 1.5, \beta = 1$. Each data point is obtained with 50,000 shots.

We noted the slight difference between the classical computing result ("EACP 1IC" in the plot) and the quantum simulator result, and verified that the error in the variational quantum algorithm is only coming from the shot error, not the numerically instability of solving the differential equation (17) (as the matrix $A$ can be singular at some time point). We plotted the error scales using box plot[66] for different number of shots, ranging from 100 to 1,000,000. For each element in matrix $A$ and vector $C$, the statistics are taken from 1,000 timesteps in evolving equation (17) with quantum circuits. The results are shown in Figure 2 (a)–(c). The horizontal orange bar indicates the median, the box encompasses 50% of the data points, and the boundary of the whisker encloses 99% of the data. It is evident that the error is solely the result of the sampling error of the measurements, and therefore confirms the algorithm's robustness in its numerical convergence.

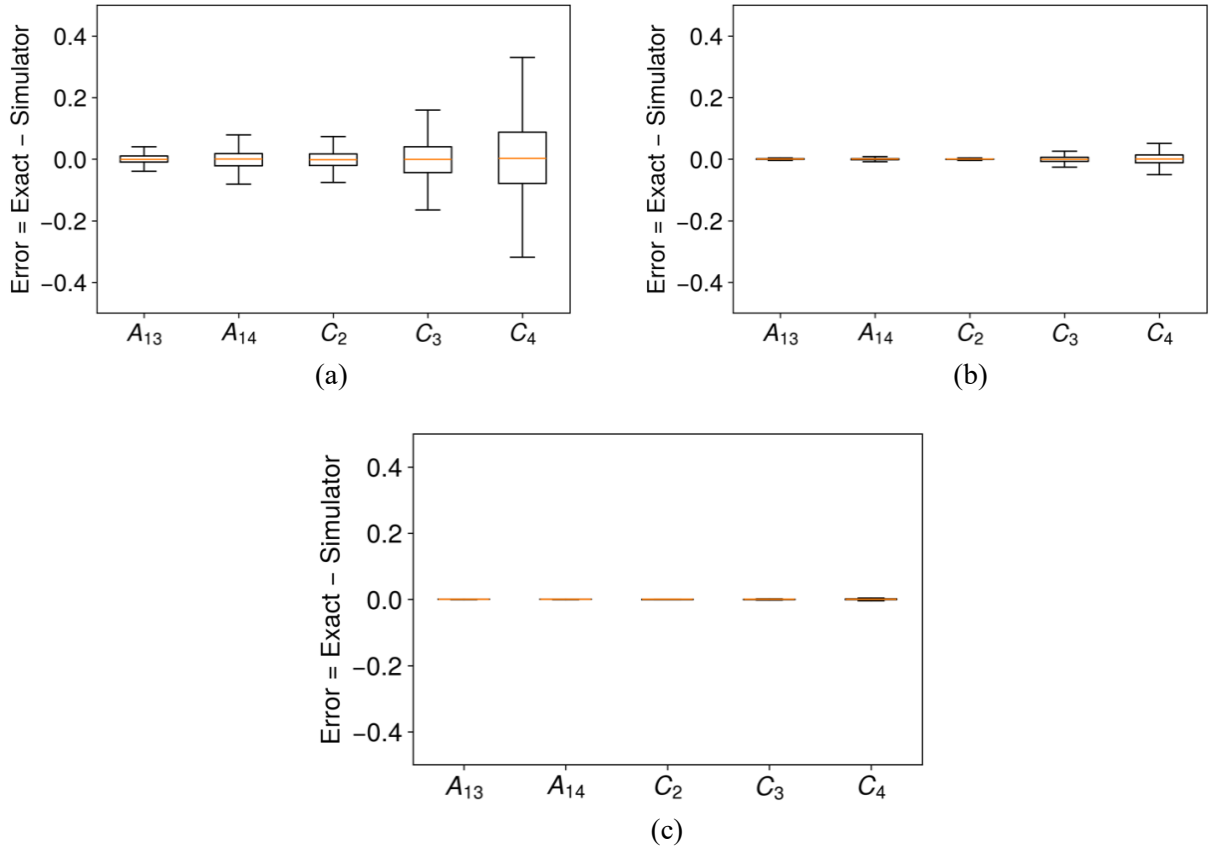

Figure 2. Shot noise comparisons for different number of shots for non-zero elements in $A$ and $C$. (a) shots = 100, (b) shots = 10,000, (c) shots = 1,000,000

In Figure 3, we present the simulation results on the *ibmq_quito* device, with the same parameters as in Figure 1. The trends show quantitative agreement. The results on the device diverge more and more from the exact result due to accumulative error in $\boldsymbol{\theta}$ from previous steps.

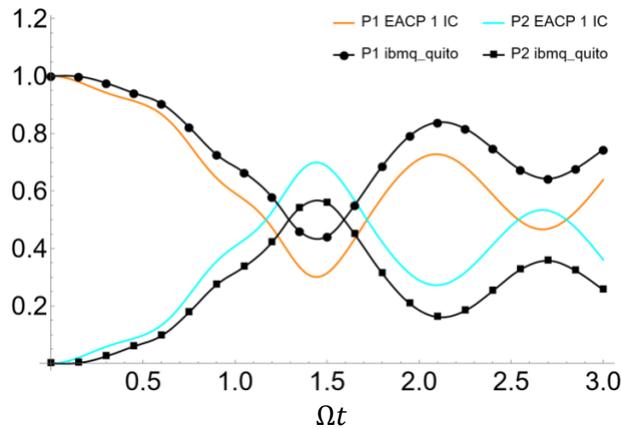

Figure 3. Population dynamics on *ibm_quito* device for a symmetric two-level system, with one bath initial condition and system initially populated in the reactant state. Parameters: $\Omega = 1, \xi = 2, \omega_c = 1.5, \beta = 1$. Each data point is obtained with 50,000 shots.

To incorporate the full thermal bath effect, Monte Carlo integration of the bath degrees of freedom needs to be performed. We conduct the analysis on the number of Monte Carlo points necessary for sampling the Wigner distribution (equation 7) to get the converged results. The findings are shown in Figure 4. It turns out the number of points needed is on the order of $10^3$ to $10^4$. Therefore, it is quite promising that with the number of qubits currently available on NISQ devices, this variational quantum algorithm can be implemented in a parallel computing fashion, with each set of qubits evolving along specific Monte Carlo points and then performing the ensemble average.

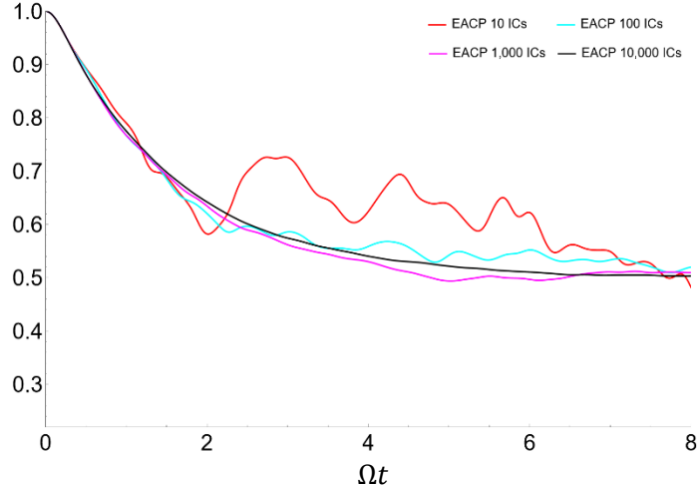

Figure 4. Monte Carlo convergence, with parameters $\Omega = 1$, $\epsilon = 0$, $\xi = 1.2$, $\omega_c = 2.5$, $\beta = 0.2$.

To test the above idea on the performance of the current quantum device, we conducted the simulations of the dynamics under the full dissipative bath (60 oscillators; 10,000 initial conditions) with the real-time noise profile from *ibm_brisbane*. The system bath parameters in these simulations are taken from reference 57. In Figure 5 (a) and (b), five curves are on display. As a reference, we include the "QuAPI" curve, which produces the numerically exact result using quasi-adiabatic propagator path integral method[67,68] with the back-reaction fully accounted for. The "EACP" curve omits the harmonic back-reaction. These two comparisons confirm the validity of using EACP as a good approximation to the exact non-Markovian quantum dynamics. The "TDVA" curve results from numerically solving equation (17). The "Simulator" simulates the variational algorithm by compiling equation (15) – (16) into quantum circuits and obtaining the measurement result. The "Noisy Simulator" simulates the variational algorithm with the real-time noise profile from the quantum device *ibm_brisbane*. Each point on the "Simulator" and "Noisy Simulator" is obtained with 50,000 shots.

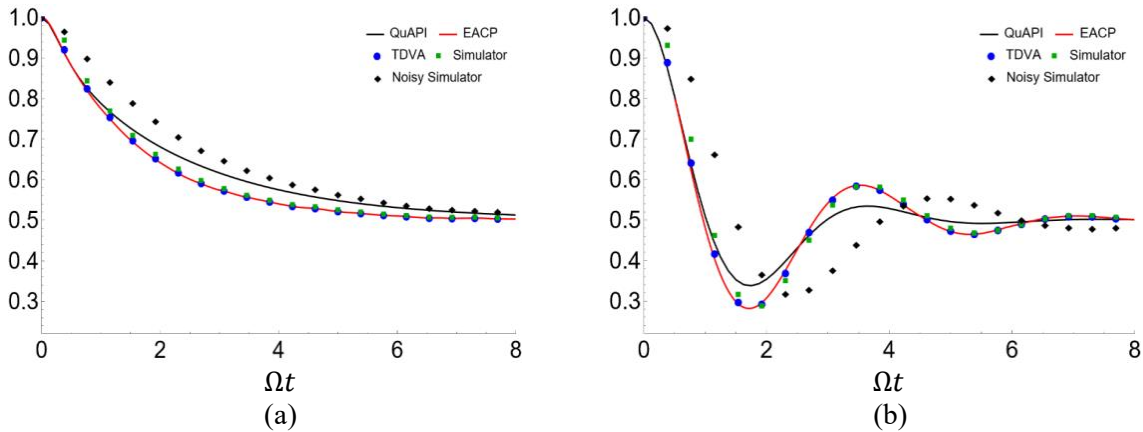

Figure 5. Population dynamics simulated for a symmetric two-level system, with 10,000 bath initial condition and system initially populated in the reactant state. Parameters for (a) $\Omega = 1$, $\epsilon = 0$, $\xi = 1.2$, $\omega_c = 2.5$, $\beta = 0.2$, and (b) $\Omega = 1$, $\epsilon = 0$, $\xi = 0.3$, $\omega_c = 5$, $\beta = 5$. Each data point on the "Simulator" and the "Noisy Simulator" curve is obtained with 50,000 shots.

To gain further insight about the ensemble averaging effect on the errors induced by the device noise, we performed simulations of population dynamics with only one bath initial condition. The representative results are shown in Figure 6 (a)–(d) and Figure 7 (a)–(d), with each graph being the dynamics from a randomly chosen initial condition. The parameters are the same as those in Figure 5. With only one initial condition, the dynamics does not dissipate. The average from all possible initial conditions produces "EACP 10,000 IC" result, plotted as a reference. The "TDVA", "Simulator" and "Noisy Simulator" results are plotted for comparison with each other. It is very apparent from the results of the "Noisy Simulator" that the device noise has different effects on different initial conditions, some diverging greatly and some bounding the accurate result to some degree, with no consistent pattern. However, when averaging them together as shown in Figure 5, it eliminates much of the randomness and the noise seems to corrupt the data with a consistent hysteresis effect. This drift in principle can be efficiently accounted for with a simple noise model.[69] Since each of these one-initial-condition dynamical simulations, equivalent to a Hamiltonian simulation, suffers from stochastic noise, whereas the statistically averaged dynamics "diversifies" away much of the random noise, it points to the advantage of using NISQ devices for simulating open quantum systems with the ensemble average approach.

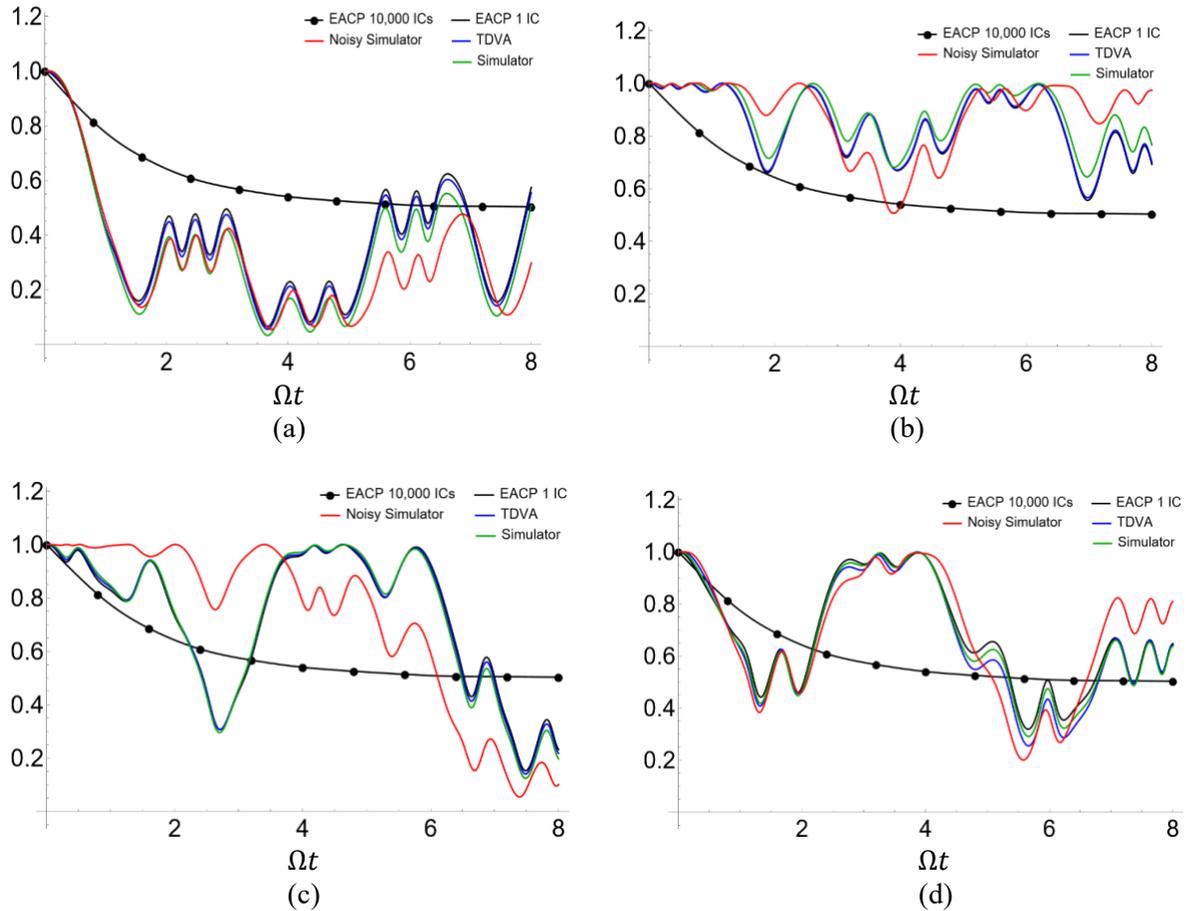

Figure 6 (a)–(d). Population dynamics simulated for a symmetric two-level system, with one bath initial condition and system initially populated in the reactant state. The parameters are $\Omega = 1$, $\epsilon = 0$, $\xi = 1.2$, $\omega_c = 2.5$, $\beta = 0.2$.

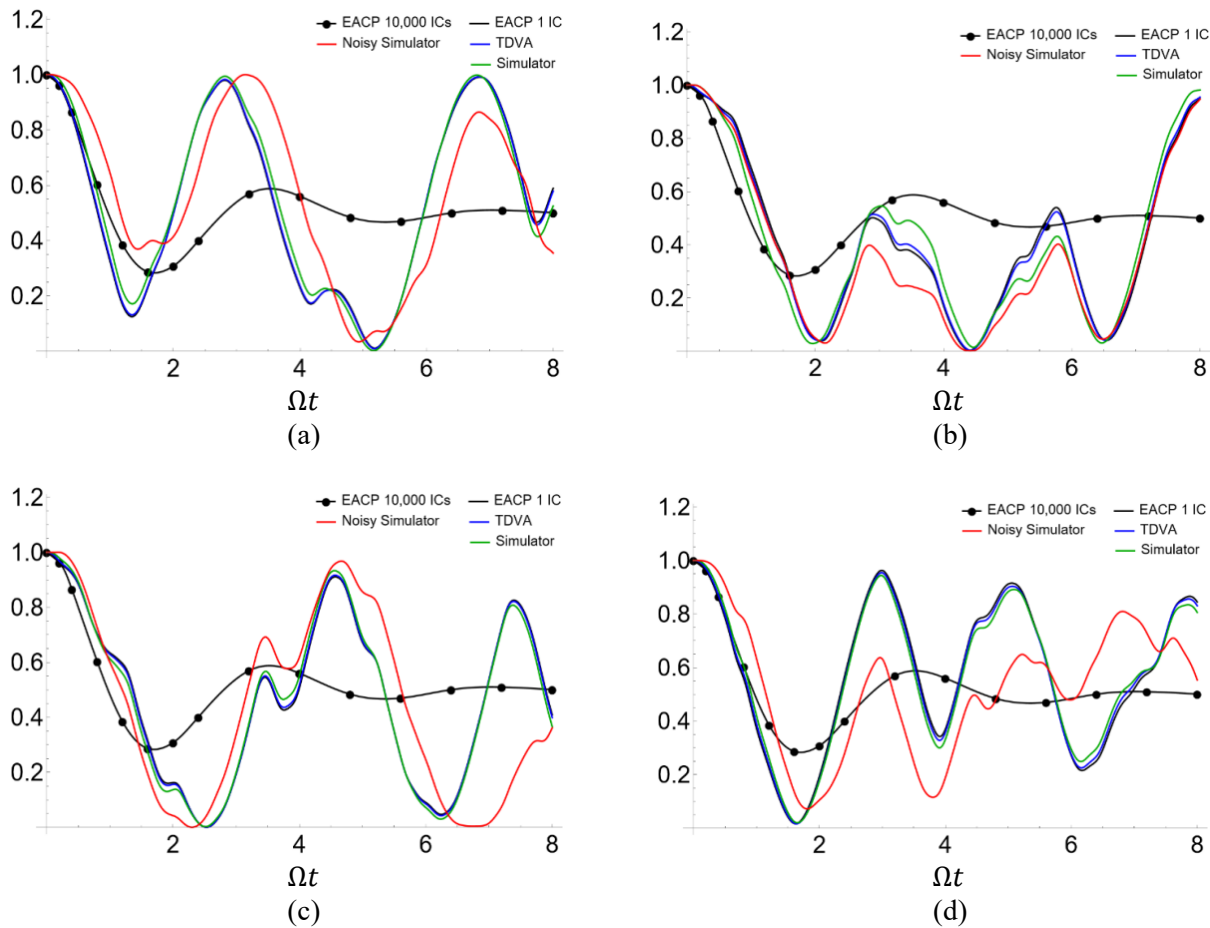

Figure 7 (a)–(d). Population dynamics simulated for a symmetric two-level system, with one bath initial condition and system initially populated in the reactant state. The parameters are $\Omega = 1$, $\epsilon = 0$, $\xi = 0.3$, $\omega_c = 5$, $\beta = 5$.

## V. Conclusions

We presented a time-dependent variational quantum algorithm based on the ensemble averaged classical path (EACP) scheme that captures much of the non-Markovian effect in quantum dynamics at finite temperature. It can become increasingly accurate as temperature increases or as the system strongly couples to the low frequency modes of the bath. We have demonstrated its feasibility on NISQ devices for the spin-boson model. The number of Monte Carlo points needed is mild, pointing to the possibility of its parallel implementation on the current NISQ devices. Furthermore, the noise effect on the dynamics is more benign compared to the Hamiltonian simulation, suggesting that open quantum dynamics simulations on NISQ devices with the ensemble average approach might be more immune to the device noise. The algorithm can be naturally extended to anharmonic bath and non-linear system-bath coupling. For its generalization to multi-site problems, such as spin chain dissipative dynamics,[70,71] a good ansatz for the time-evolution operator [53,54,72–74] is crucial for avoiding exponential time compilation and measurement overhead.

**Acknowledgement**
This work is supported by National Science Foundation (NSF) under Award 2320328, and George Mason University's startup fund and its Science and Engineering Center (QSEC) funding. This work used Explore ACCESS[75] at SDSC Expanse CPU through allocation CHE220009 from the Advanced Cyberinfrastructure Coordination Ecosystem: Services & Support (ACCESS) program, which is supported by National Science Foundation grants 2138259, 2138286, 2138307, 2137603, and 2138296. We also acknowledge the use of IBM Quantum services for this work. The views expressed are those of the authors and do not reflect the official policy or position of IBM or the IBM Quantum team.

**Appendix**

The following shows exemplar circuit for matrix *A* and vector *C*.

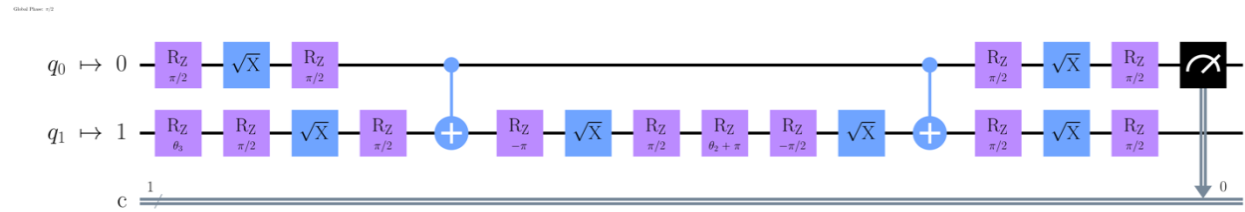

Figure A.1. $A_{13}$ circuit.

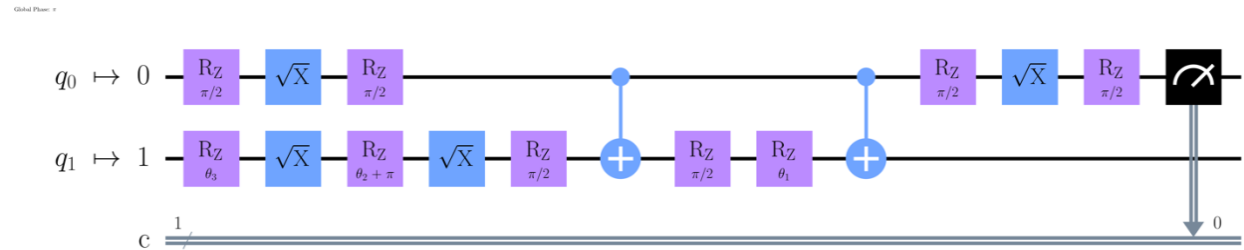

Figure A.2. $C_2$ circuit.

**Reference**
(1) Nitzan, A. *Chemical Dynamics in Condensed Phases: Relaxation, Transfer and Reactions in Condensed Molecular Systems*, 1. publ. in paperback, [corr. ed.].; Oxford graduate texts; Oxford University Press: Oxford, 2013.
(2) Breuer, H.-P.; Petruccione, F. *The Theory of Open Quantum Systems*, Repr.; Clarendon Press: Oxford, 2010.
(3) Mohseni, M.; Omar, Y.; Engel, G.; Plenio, M. B. *Quantum Effects in Biology*; Cambridge University Press: Cambridge, 2014.
(4) Nielsen, M. A.; Chuang, I. L. *Quantum Computation and Quantum Information: 10th Anniversary Edition*, 1st ed.; Cambridge University Press, 2012. https://doi.org/10.1017/CBO9780511976667.
(5) Jo, M.; Um, J.; Kahng, B. Nonequilibrium Phase Transition in an Open Quantum Spin System with Long-Range Interaction. *Phys. Rev. E* **2019**, *99* (3), 032131. https://doi.org/10.1103/PhysRevE.99.032131.
(6) Pižorn, I. One-Dimensional Bose-Hubbard Model Far from Equilibrium. *Phys. Rev. A* **2013**, *88* (4), 043635. https://doi.org/10.1103/PhysRevA.88.043635.
(7) Kraus, B.; Büchler, H. P.; Diehl, S.; Kantian, A.; Micheli, A.; Zoller, P. Preparation of Entangled States by Quantum Markov Processes. *Phys. Rev. A* **2008**, *78* (4), 042307. https://doi.org/10.1103/PhysRevA.78.042307.

(8) Hartmann, L.; Dür, W.; Briegel, H.-J. Steady-State Entanglement in Open and Noisy Quantum Systems. *Phys. Rev. A* **2006**, *74* (5), 052304. https://doi.org/10.1103/PhysRevA.74.052304.

(9) Breuer, H.-P.; Laine, E.-M.; Piilo, J. Measure for the Degree of Non-Markovian Behavior of Quantum Processes in Open Systems. *Phys. Rev. Lett.* **2009**, *103* (21), 210401. https://doi.org/10.1103/PhysRevLett.103.210401.

(10) Lu, X.-M.; Wang, X.; Sun, C. P. Quantum Fisher Information Flow and Non-Markovian Processes of Open Systems. *Phys. Rev. A* **2010**, *82* (4), 042103. https://doi.org/10.1103/PhysRevA.82.042103.

(11) Lorenzo, S.; Plastina, F.; Paternostro, M. Geometrical Characterization of Non-Markovianity. *Phys. Rev. A* **2013**, *88* (2), 020102. https://doi.org/10.1103/PhysRevA.88.020102.

(12) Walters, P. L.; Makri, N. Quantum–Classical Path Integral Simulation of Ferrocene–Ferrocenium Charge Transfer in Liquid Hexane. *J. Phys. Chem. Lett.* **2015**, *6* (24), 4959–4965. https://doi.org/10.1021/acs.jpclett.5b02265.

(13) Warshel, A.; Parson, W. W. Computer Simulations of Electron-Transfer Reactions in Solution and in Photosynthetic Reaction Centers. *Annu. Rev. Phys. Chem.* **1991**, *42* (1), 279–309. https://doi.org/10.1146/annurev.pc.42.100191.001431.

(14) Kundu, S.; Dani, R.; Makri, N. Tight Inner Ring Architecture and Quantum Motion of Nuclei Enable Efficient Energy Transfer in Bacterial Light Harvesting. *Sci. Adv.* **2022**, *8* (43), eadd0023. https://doi.org/10.1126/sciadv.add0023.

(15) Kundu, S.; Makri, N. Exciton–Vibration Dynamics in J-Aggregates of a Perylene Bisimide from Real-Time Path Integral Calculations. *J. Phys. Chem. C* **2021**, *125* (1), 201–210. https://doi.org/10.1021/acs.jpcc.0c09405.

(16) Fay, T. P.; Limmer, D. T. Coupled Charge and Energy Transfer Dynamics in Light Harvesting Complexes from a Hybrid Hierarchical Equations of Motion Approach. *J. Chem. Phys.* **2022**, *157* (17), 174104. https://doi.org/10.1063/5.0117659.

(17) Binder, R.; Lauvergnat, D.; Burghardt, I. Conformational Dynamics Guides Coherent Exciton Migration in Conjugated Polymer Materials: First-Principles Quantum Dynamical Study. *Phys. Rev. Lett.* **2018**, *120* (22), 227401. https://doi.org/10.1103/PhysRevLett.120.227401.

(18) Leggett, A. J.; Chakravarty, S.; Dorsey, A. T.; Fisher, M. P. A.; Garg, A.; Zwerger, W. Dynamics of the Dissipative Two-State System. *Rev. Mod. Phys.* **1987**, *59* (1), 1–85. https://doi.org/10.1103/RevModPhys.59.1.

(19) Weiss, U. *Quantum Dissipative Systems*, 4th ed.; WORLD SCIENTIFIC, 2012. https://doi.org/10.1142/8334.

(20) Makri, N. Small Matrix Path Integral for System-Bath Dynamics. *J. Chem. Theory Comput.* **2020**, *16* (7), 4038–4049. https://doi.org/10.1021/acs.jctc.0c00039.

(21) Barford, W.; Tozer, O. R. Theory of Exciton Transfer and Diffusion in Conjugated Polymers. *J. Chem. Phys.* **2014**, *141* (16), 164103. https://doi.org/10.1063/1.4897986.

(22) Topaler, M.; Makri, N. Quasi-Adiabatic Propagator Path Integral Methods. Exact Quantum Rate Constants for Condensed Phase Reactions. *Chem. Phys. Lett.* **1993**, *210* (1–3), 285–293. https://doi.org/10.1016/0009-2614(93)89135-5.

(23) Tanimura, Y. Numerically "Exact" Approach to Open Quantum Dynamics: The Hierarchical Equations of Motion (HEOM). *J. Chem. Phys.* **2020**, *153* (2), 020901. https://doi.org/10.1063/5.0011599.

(24) Beck, M. The Multiconfiguration Time-Dependent Hartree (MCTDH) Method: A Highly Efficient Algorithm for Propagating Wavepackets. *Phys. Rep.* **2000**, *324* (1), 1–105. https://doi.org/10.1016/S0370-1573(99)00047-2.

(25) Makri, N. Small Matrix Disentanglement of the Path Integral: Overcoming the Exponential Tensor Scaling with Memory Length. *J. Chem. Phys.* **2020**, *152* (4), 041104. https://doi.org/10.1063/1.5139473.

(26) Lambert, R.; Makri, N. Quantum-Classical Path Integral. II. Numerical Methodology. *J. Chem. Phys.* **2012**, *137* (22), 22A553. https://doi.org/10.1063/1.4767980.

(27) Bose, A.; Walters, P. L. A Multisite Decomposition of the Tensor Network Path Integrals. *J. Chem. Phys.* **2022**, *156* (2), 024101. https://doi.org/10.1063/5.0073234.

(28) Feynman, R. P. Simulating Physics with Computers. *Int. J. Theor. Phys.* **1982**, *21* (6–7), 467–488. https://doi.org/10.1007/BF02650179.

(29) Lloyd, S. Universal Quantum Simulators. *Science* **1996**, *273* (5278), 1073–1078. https://doi.org/10.1126/science.273.5278.1073.

(30) Berry, D. W.; Childs, A. M.; Cleve, R.; Kothari, R.; Somma, R. D. Simulating Hamiltonian Dynamics with a Truncated Taylor Series. *Phys. Rev. Lett.* **2015**, *114* (9), 090502. https://doi.org/10.1103/PhysRevLett.114.090502.

(31) Berry, D. W.; Ahokas, G.; Cleve, R.; Sanders, B. C. Efficient Quantum Algorithms for Simulating Sparse Hamiltonians. *Commun. Math. Phys.* **2007**, *270* (2), 359–371. https://doi.org/10.1007/s00220-006-0150-x.

(32) Childs, A. M.; Maslov, D.; Nam, Y.; Ross, N. J.; Su, Y. Toward the First Quantum Simulation with Quantum Speedup. *Proc. Natl. Acad. Sci.* **2018**, *115* (38), 9456–9461. https://doi.org/10.1073/pnas.1801723115.

(33) Campbell, E. Random Compiler for Fast Hamiltonian Simulation. *Phys. Rev. Lett.* **2019**, *123* (7), 070503. https://doi.org/10.1103/PhysRevLett.123.070503.

(34) Cîrstoiu, C.; Holmes, Z.; Iosue, J.; Cincio, L.; Coles, P. J.; Sornborger, A. Variational Fast Forwarding for Quantum Simulation beyond the Coherence Time. *Npj Quantum Inf.* **2020**, *6* (1), 82. https://doi.org/10.1038/s41534-020-00302-0.

(35) Gilyén, A.; Su, Y.; Low, G. H.; Wiebe, N. Quantum Singular Value Transformation and beyond: Exponential Improvements for Quantum Matrix Arithmetics. In *Proceedings of the 51st Annual ACM SIGACT Symposium on Theory of Computing*; ACM: Phoenix AZ USA, 2019; pp 193–204. https://doi.org/10.1145/3313276.3316366.

(36) Low, G. H.; Chuang, I. L. Optimal Hamiltonian Simulation by Quantum Signal Processing. *Phys. Rev. Lett.* **2017**, *118* (1), 010501. https://doi.org/10.1103/PhysRevLett.118.010501.

(37) Lindblad, G. On the Generators of Quantum Dynamical Semigroups. *Commun. Math. Phys.* **1976**, *48* (2), 119–130. https://doi.org/10.1007/BF01608499.

(38) Bacon, D.; Childs, A. M.; Chuang, I. L.; Kempe, J.; Leung, D. W.; Zhou, X. Universal Simulation of Markovian Quantum Dynamics. *Phys. Rev. A* **2001**, *64* (6), 062302. https://doi.org/10.1103/PhysRevA.64.062302.

(39) Lidar, D. A.; Bihary, Z.; Whaley, K. B. From Completely Positive Maps to the Quantum Markovian Semigroup Master Equation. *Chem. Phys.* **2001**, *268* (1–3), 35–53. https://doi.org/10.1016/S0301-0104(01)00330-5.

(40) Childs, A. M.; Li, T. Efficient Simulation of Sparse Markovian Quantum Dynamics. **2016**. https://doi.org/10.48550/ARXIV.1611.05543.

(41) Sweke, R.; Sinayskiy, I.; Bernard, D.; Petruccione, F. Universal Simulation of Markovian Open Quantum Systems. *Phys. Rev. A* **2015**, *91* (6), 062308. https://doi.org/10.1103/PhysRevA.91.062308.

(42) Hu, Z.; Xia, R.; Kais, S. A Quantum Algorithm for Evolving Open Quantum Dynamics on Quantum Computing Devices. *Sci. Rep.* **2020**, *10* (1), 3301. https://doi.org/10.1038/s41598-020-60321-x.

(43) Schlimgen, A. W.; Head-Marsden, K.; Sager, L. M.; Narang, P.; Mazziotti, D. A. Quantum Simulation of Open Quantum Systems Using a Unitary Decomposition of Operators. *Phys. Rev. Lett.* **2021**, *127* (27), 270503. https://doi.org/10.1103/PhysRevLett.127.270503.

(44) Schlimgen, A. W.; Head-Marsden, K.; Sager, L. M.; Narang, P.; Mazziotti, D. A. Quantum Simulation of the Lindblad Equation Using a Unitary Decomposition of Operators. *Phys. Rev. Res.* **2022**, *4* (2), 023216. https://doi.org/10.1103/PhysRevResearch.4.023216.

(45) Schlimgen, A. W.; Head-Marsden, K.; Sager-Smith, L. M.; Narang, P.; Mazziotti, D. A. Quantum State Preparation and Nonunitary Evolution with Diagonal Operators. *Phys. Rev. A* **2022**, *106* (2), 022414. https://doi.org/10.1103/PhysRevA.106.022414.

(46) Hu, Z.; Head-Marsden, K.; Mazziotti, D. A.; Narang, P.; Kais, S. A General Quantum Algorithm for Open Quantum Dynamics Demonstrated with the Fenna-Matthews-Olson Complex. *Quantum* **2022**, *6*, 726. https://doi.org/10.22331/q-2022-05-30-726.
(47) Sweke, R.; Sanz, M.; Sinayskiy, I.; Petruccione, F.; Solano, E. Digital Quantum Simulation of Many-Body Non-Markovian Dynamics. *Phys. Rev. A* **2016**, *94* (2), 022317. https://doi.org/10.1103/PhysRevA.94.022317.
(48) Head-Marsden, K.; Krastanov, S.; Mazziotti, D. A.; Narang, P. Capturing Non-Markovian Dynamics on near-Term Quantum Computers. *Phys. Rev. Res.* **2021**, *3* (1), 013182. https://doi.org/10.1103/PhysRevResearch.3.013182.
(49) Wang, Y.; Mulvihill, E.; Hu, Z.; Lyu, N.; Shivpuje, S.; Liu, Y.; Soley, M. B.; Geva, E.; Batista, V. S.; Kais, S. Simulating Open Quantum System Dynamics on NISQ Computers with Generalized Quantum Master Equations. *J. Chem. Theory Comput.* **2023**, *19* (15), 4851–4862. https://doi.org/10.1021/acs.jctc.3c00316.
(50) Seneviratne, A.; Walters, P. L.; Wang, F. Exact Non-Markovian Quantum Dynamics on the NISQ Device Using Kraus Operators. *ACS Omega* **2024**, acsomega.3c09720. https://doi.org/10.1021/acsomega.3c09720.
(51) Walters, P. L.; Wang, F. Path Integral Quantum Algorithm for Simulating Non-Markovian Quantum Dynamics in Open Quantum Systems. *Phys. Rev. Res.* **2024**, *6* (1), 013135. https://doi.org/10.1103/PhysRevResearch.6.013135.
(52) Endo, S.; Sun, J.; Li, Y.; Benjamin, S. C.; Yuan, X. Variational Quantum Simulation of General Processes. *Phys. Rev. Lett.* **2020**, *125* (1), 010501. https://doi.org/10.1103/PhysRevLett.125.010501.
(53) Cerezo, M.; Arrasmith, A.; Babbush, R.; Benjamin, S. C.; Endo, S.; Fujii, K.; McClean, J. R.; Mitarai, K.; Yuan, X.; Cincio, L.; Coles, P. J. Variational Quantum Algorithms. *Nat. Rev. Phys.* **2021**, *3* (9), 625–644. https://doi.org/10.1038/s42254-021-00348-9.
(54) Benedetti, M.; Fiorentini, M.; Lubasch, M. Hardware-Efficient Variational Quantum Algorithms for Time Evolution. *Phys. Rev. Res.* **2021**, *3* (3), 033083. https://doi.org/10.1103/PhysRevResearch.3.033083.
(55) Makri, N. Exploiting Classical Decoherence in Dissipative Quantum Dynamics: Memory, Phonon Emission, and the Blip Sum. *Chem. Phys. Lett.* **2014**, *593*, 93–103. https://doi.org/10.1016/j.cplett.2013.11.064.
(56) Banerjee, T.; Makri, N. Quantum-Classical Path Integral with Self-Consistent Solvent-Driven Reference Propagators. *J. Phys. Chem. B* **2013**, *117* (42), 13357–13366. https://doi.org/10.1021/jp4043123.
(57) Lambert, R.; Makri, N. Quantum-Classical Path Integral. I. Classical Memory and Weak Quantum Nonlocality. *J. Chem. Phys.* **2012**, *137* (22), 22A552. https://doi.org/10.1063/1.4767931.
(58) Wang, F.; Makri, N. Quantum-Classical Path Integral with a Harmonic Treatment of the Back-Reaction. *J. Chem. Phys.* **2019**, *150* (18), 184102. https://doi.org/10.1063/1.5091725.
(59) Feynman, R. P.; Vernon, F. L. The Theory of a General Quantum System Interacting with a Linear Dissipative System. *Ann. Phys.* **1963**, *24*, 118–173. https://doi.org/10.1016/0003-4916(63)90068-X.
(60) Hastings, W. K. Monte Carlo Sampling Methods Using Markov Chains and Their Applications. *Biometrika* **1970**, *57* (1), 97–109. https://doi.org/10.1093/biomet/57.1.97.
(61) Bose, A.; Makri, N. Wigner Phase Space Distribution via Classical Adiabatic Switching. *J. Chem. Phys.* **2015**, *143* (11), 114114. https://doi.org/10.1063/1.4930271.
(62) Yuan, X.; Endo, S.; Zhao, Q.; Li, Y.; Benjamin, S. C. Theory of Variational Quantum Simulation. *Quantum* **2019**, *3*, 191. https://doi.org/10.22331/q-2019-10-07-191.
(63) Li, Y.; Benjamin, S. C. Efficient Variational Quantum Simulator Incorporating Active Error Minimization. *Phys. Rev. X* **2017**, *7* (2), 021050. https://doi.org/10.1103/PhysRevX.7.021050.
(64) Walters, P. L.; Allen, T. C.; Makri, N. Direct Determination of Discrete Harmonic Bath Parameters from Molecular Dynamics Simulations. *J. Comput. Chem.* **2017**, *38* (2), 110–115. https://doi.org/10.1002/jcc.24527.

(65) Aleksandrowicz, G.; Alexander, T.; Barkoutsos, P.; Bello, L.; Ben-Haim, Y.; Bucher, D.; Cabrera-Hernández, F. J.; Carballo-Franquis, J.; Chen, A.; Chen, C.-F.; Chow, J. M.; Córcoles-Gonzales, A. D.; Cross, A. J.; Cross, A.; Cruz-Benito, J.; Culver, C.; González, S. D. L. P.; Torre, E. D. L.; Ding, D.; Dumitrescu, E.; Duran, I.; Eendebak, P.; Everitt, M.; Sertage, I. F.; Frisch, A.; Fuhrer, A.; Gambetta, J.; Gago, B. G.; Gomez-Mosquera, J.; Greenberg, D.; Hamamura, I.; Havlicek, V.; Hellmers, J.; Łukasz Herok; Horii, H.; Shaohan Hu; Imamichi, T.; Toshinari Itoko; Javadi-Abhari, A.; Kanazawa, N.; Karazeev, A.; Krsulich, K.; Liu, P.; Luh, Y.; Yunho Maeng; Marques, M.; Martín-Fernández, F. J.; McClure, D. T.; McKay, D.; Srujan Meesala; Mezzacapo, A.; Moll, N.; Rodríguez, D. M.; Nannicini, G.; Nation, P.; Ollitrault, P.; O'Riordan, L. J.; Hanhee Paik; Pérez, J.; Phan, A.; Pistoia, M.; Prutyanov, V.; Reuter, M.; Rice, J.; Abdón Rodríguez Davila; Rudy, R. H. P.; Mingi Ryu; Ninad Sathaye; Schnabel, C.; Schoute, E.; Kanav Setia; Yunong Shi; Adenilton Silva; Siraichi, Y.; Seyon Sivarajah; Smolin, J. A.; Soeken, M.; Takahashi, H.; Tavernelli, I.; Taylor, C.; Taylour, P.; Kenso Trabing; Treinish, M.; Turner, W.; Vogt-Lee, D.; Vuillot, C.; Wildstrom, J. A.; Wilson, J.; Winston, E.; Wood, C.; Wood, S.; Wörner, S.; Akhalwaya, I. Y.; Zoufal, C. Qiskit: An Open-Source Framework for Quantum Computing, 2019. https://doi.org/10.5281/ZENODO.2562110.

(66) Dutoit, S. H. C. *Graphical Exploratory Data Analysis*; Springer: Place of publication not identified, 2012.

(67) Makri, N. Improved Feynman Propagators on a Grid and Non-Adiabatic Corrections within the Path Integral Framework. *Chem. Phys. Lett.* **1992**, *193* (5), 435–445. https://doi.org/10.1016/0009-2614(92)85654-S.

(68) Topaler, M.; Makri, N. System-Specific Discrete Variable Representations for Path Integral Calculations with Quasi-Adiabatic Propagators. *Chem. Phys. Lett.* **1993**, *210* (4–6), 448–457. https://doi.org/10.1016/0009-2614(93)87052-5.

(69) Velázquez, J. M. S.; Steiner, A.; Freund, R.; Guevara-Bertsch, M.; Marciniak, Ch. D.; Monz, T.; Bermudez, A. Dynamical Quantum Maps for Single-Qubit Gates under Non-Markovian Phase Noise. **2024**. https://doi.org/10.48550/ARXIV.2402.14530.

(70) Sadiek, G.; Almalki, S. Entanglement Dynamics in Heisenberg Spin Chains Coupled to a Dissipative Environment at Finite Temperature. *Phys. Rev. A* **2016**, *94* (1), 012341. https://doi.org/10.1103/PhysRevA.94.012341.

(71) Hebenstreit, F.; Banerjee, D.; Hornung, M.; Jiang, F.-J.; Schranz, F.; Wiese, U.-J. Real-Time Dynamics of Open Quantum Spin Systems Driven by Dissipative Processes. *Phys. Rev. B* **2015**, *92* (3), 035116. https://doi.org/10.1103/PhysRevB.92.035116.

(72) Grimsley, H. R.; Economou, S. E.; Barnes, E.; Mayhall, N. J. An Adaptive Variational Algorithm for Exact Molecular Simulations on a Quantum Computer. *Nat. Commun.* **2019**, *10* (1), 3007. https://doi.org/10.1038/s41467-019-10988-2.

(73) Yao, Y.-X.; Gomes, N.; Zhang, F.; Wang, C.-Z.; Ho, K.-M.; Iadecola, T.; Orth, P. P. Adaptive Variational Quantum Dynamics Simulations. *PRX Quantum* **2021**, *2* (3), 030307. https://doi.org/10.1103/PRXQuantum.2.030307.

(74) Choquette, A.; Di Paolo, A.; Barkoutsos, P. Kl.; Sénéchal, D.; Tavernelli, I.; Blais, A. Quantum-Optimal-Control-Inspired Ansatz for Variational Quantum Algorithms. *Phys. Rev. Res.* **2021**, *3* (2), 023092. https://doi.org/10.1103/PhysRevResearch.3.023092.

(75) Boerner, T. J.; Deems, S.; Furlani, T. R.; Knuth, S. L.; Towns, J. ACCESS: Advancing Innovation: NSF's Advanced Cyberinfrastructure Coordination Ecosystem: Services & Support. In *Practice and Experience in Advanced Research Computing*; ACM: Portland OR USA, 2023; pp 173–176. https://doi.org/10.1145/3569951.3597559.